\documentclass[preprint]{aastex}

\begin{document}

\title {Physical Properties of the Low-Mass Eclipsing Binary NSVS 02502726}
\author{Jae Woo Lee, Jae-Hyuck Youn, Seung-Lee Kim, and Chung-Uk Lee}
\affil{Korea Astronomy and Space Science Institute, Daejon 305-348, Korea}
\email{jwlee@kasi.re.kr, jhyoon@kasi.re.kr, slkim@kasi.re.kr, leecu@kasi.re.kr}

\begin{abstract}
NSVS 02502726 has been known as a double-lined, detached eclipsing binary that consists of two low-mass stars. We obtained $BVRI$ 
photometric follow-up observations in 2009 and 2011 to measure improved physical properties of the binary star. Each set of light curves, 
including the 2008 data given by \c Cakirli et al., was simultaneously analyzed with the previously published radial-velocity curves 
using the Wilson-Devinney binary code. The conspicuous seasonal light variations of the system are satisfactorily modelled by 
a two-spot model with one starspot on each component and by changes of the spot parameters with time. Based on 23 eclipse timings 
calculated from the synthetic model and one ephemeris epoch, an orbital period study of NSVS 02502726 reveals that the period
has experienced a continuous decrease of $-$5.9$\times$10$^{-7}$ d yr$^{-1}$ or a sinusoidal variation with a period and 
semi-amplitude of 2.51 yrs and 0.0011 d, respectively. The timing variations could be interpreted as either 
the light-travel-time effect due to the presence of an unseen third body, or as the combination of this effect and angular momentum loss 
via magnetic stellar wind braking. Individual masses and radii of both components are determined to be $M_1$=0.689$\pm$0.016 M$_\odot$, 
$M_2$=0.341$\pm$0.009 M$_\odot$, $R_1$=0.707$\pm$0.007 R$_\odot$, and $R_2$=0.657$\pm$0.008 R$_\odot$. The results are very different 
from those of \c Cakirli et al. with the primary's radius (0.674$\pm$0.006 R$_\odot$) smaller the secondary's (0.763$\pm$0.007 R$_\odot$). 
We compared the physical parameters presented in this paper with current low-mass stellar models and found that the measured values 
of the primary star are best fitted to a 79-Myr isochrone. The primary is in good agreement with the empirical mass-radius relation 
from low-mass binaries, but the secondary is oversized by about 85\%.
\end{abstract}

{}

\section{INTRODUCTION}

The study of detached, double-lined eclipsing binaries provides an accurate and direct determination of fundamental stellar properties 
such as mass, radius, and luminosity (Hilditch 2001). Because their uncertainties are below $\sim$ 1\% in the best-observed systems, 
these data can allow us to test stellar structure and evolution models and to calculate the distance to the eclipsing system
(Andersen 1991; Torres et al. 2010). Generally, the results of theoretical stellar models compare favorably to the observed properties 
for main-sequence stars with masses greater than the Sun, while the current models for lower main sequence stars have not yet matched 
well with observations. The radii and temperatures of the low-mass stars are often observed to be $\sim$ 10\% larger and $\sim$ 5\% cooler, 
respectively, than predicted by models for their masses (Ribas 2006; Morales et al. 2010). Most recently, Feiden \& Chaboyer (2012) 
shows that the radius deviations typically less than 4\%, by comparing the low-mass models of the Dartmouth stellar evolution database 
(Dotter et al. 2008) with detached eclipsing binary systems. The discrepancies are mainly caused by the effect of magnetic activity, 
which produces outstanding surface activity (spots) in the binary systems with both short orbital periods and deep convective envelopes 
(L\'opez-Morales \& Ribas 2005; Chabrier et al. 2007). Morales et al. (2010) suggested that a significant spot areal coverage induces 
systematic deviations in the radii and effective temperatures derived from the light curves and that concentration of spots near 
the poles could explain the radius discrepancy between models and observations. 

In order to advance this subject, we have been observing candidate low-mass eclipsing binaries (e.g. Koo et al. 2012). 
The present paper is concerned with NSVS 02502726 (2MASS J08441103+5423473; $V$=$+$13.41, $B$--$V$=$+$1.26), which was identified 
by Coughlin \& Shaw (2007) to be a detached eclipsing pair with a short orbital period of 0.559772 d. 
\c Cakirli et al. (2009, hereafter CIG) obtained the $RI$ CCD light curves and double-lined radial-velocity data and analyzed 
these separately. Their results showed that NSVS 02502726 is a low-mass eclipsing binary, whose secondary has a larger radius than 
the primary component. The galactic space velocities of ($U, V, W$) indicated that the system is not an old object. 
Using solar metallicity models, they also reported that the component stars are in the final stages of pre-main-sequence contraction 
with a young age of 126 Myr. The aims of this paper are to present new multiband CCD photometry of NSVS 02502726 observed in both 2009 
and 2011 and to measure the physical properties of the system from detailed studies of all available data, such as radial velocities, 
light curves, and eclipse timings.

\section{NEW CCD PHOTOMETRY}

We carried out CCD photometric observations of NSVS 02502726 during two observing seasons, using $BVRI$ filters attached to the 1.0-m 
reflector at the Mt. Lemmon Optical Astronomy Observatory (LOAO) in Arizona, USA. The observations of the first season were made on 
14 nights from 2009 March to May using an FLI IMG4301E CCD camera and those of the second season were made on 16 nights from 2011 March 
to April using an ARC 4K CCD camera. The instruments and reduction methods were the same as those described by Lee et al. (2009b, 2012)
in the same order. The comparison (C) and check (K) stars were chosen to be GSC 3798-1250 (2MASS J08440529+5422332) and GSC 3798-1240.
No light variability was reported in the previous observations of these two stars by CIG and we also observed no intrinsic variability. 
The reference stars were imaged on the chip at the same time as the eclipsing variable.

From the LOAO observations, a total of 4753 individual points (1428 in 2009, 3325 in 2011) were obtained in the four bandpasses 
(1182 in $B$, 1192 in $V$, 1193 in $R$, and 1186 in $I$) and a sample of them is listed in Table 1. The 2009 and 2011 light curves of 
NSVS 02502726 are plotted in Figures 1 and 2, respectively, as differential magnitudes {\it versus} orbital phase, which was computed 
according to the ephemeris for our two-spot model described in the following section. The $B$ light curve is the first ever compiled. 
The 1$\sigma$-values of the dispersion of the (K$-$C) differences are $\pm$0.024 mag, $\pm$0.016 mag, $\pm$0.015 mag, and 
$\pm$0.006 mag from $B$ to $I$ bandpasses, respectively, for the 2009 season and $\pm$0.009 mag, $\pm$0.005 mag, $\pm$0.005 mag, and 
$\pm$0.005 mag for the 2011 season.

\section{ANALYSIS}

\subsection{Light and Velocity Curves}

As shown in Figures 1 and 2, the light curves of NSVS 02502726 were completely covered and the different depths between the primary 
and secondary eclipses indicate a large temperature difference between the two components. The LOAO light maxima (Max I and Max II) 
lie around orbital phases 0.42 and 0.58 just before and after the secondary eclipse, while those of CIG's light curves given in 
Figure 3 are displaced to around phases 0.08 and 0.92 just before and after the primary eclipse. Such phenomena have been reported 
commonly for light curves of low-mass close binaries and are usually interpreted as spot activity on component stars 
(e.g., CU Cnc, Ribas 2003; 2MASS J05162881+2607387, Bayless \& Orosz 2006; GU Boo, Windmiller et al. 2010).

In order to obtain a consistent set of the binary parameters, we analyzed all available radial-velocity and light curves of 
NSVS 02502726 by using the 2003 version of the Wilson-Devinney synthesis code (Wilson \& Devinney 1971, hereafter W-D). 
The effective temperature of the hotter, more massive primary star was initialized to be $T_{1}$=4300 K given by CIG. 
The gravity-darkening exponents and the bolometric albedos were fixed at standard values ($g$=0.32 and $A$=0.5) for stars 
with convective envelopes. The square root bolometric ($X$, $Y$) and monochromatic ($x$, $y$) limb-darkening coefficients were 
interpolated from the values of van Hamme (1993) in concert with the model atmosphere option. Furthermore, a synchronous rotation 
for both components and a circular orbit were adopted and the detailed reflection effect was considered. In this paper, 
the subscripts 1 and 2 refer to the primary and secondary stars being eclipsed at Min I (at phase 0.0) and Min II, respectively.

For the simultaneous analysis of velocity and light curves, we used a weighting scheme similar to that for the eclipsing binary
RU UMi (Lee et al. 2008). Table 2 lists the radial velocity and light-curve sets of NSVS 02502726 analyzed in this paper and 
their standard deviations ($\sigma$). Our analyses have been carried out through two stages. In the first stage, all velocity 
and light curves were simultaneously solved without spots. The result for this analysis was plotted as the dashed curves 
in Figures 1--3, where the model light curves do not fit the observed ones at all well. In the second stage, each set of 
light curves was simultaneously modelled with the radial velocity data by using the unspotted solution as the initial values 
and then including the starspot on the binary components. Final results are listed in Table 3 for the three datasets. 
The synthetic light curves are displayed as the solid curves in Figures 1--3, while the synthetic radial velocity curves are 
plotted in Figure 4. The intrinsic light variations of NSVS 02502726 have been modeled by the simultaneous existence of 
both a polar spot on the primary component and an equatorial spot on the secondary, and by the variability of the spot parameters 
with time. Our solutions represent the system as a detached binary in which both stars lie well inside 
their Roche lobes. Moreover, because the binary components should have deep outer convective layers or be fully convective as 
surmised from their temperatures and masses, it is reasonable to regard the main cause of the spot activity as a magnetic dynamo. 
In all procedures that have been described, we looked for a possible third light source ($\ell_{3}$) but found that the parameter 
remained zero within its error.
 
\subsection{Eclipse Timings}

In the low-mass close binaries displaying enhanced magnetic activity, times of minimum light are shifted from the real conjunctions 
by asymmetrical eclipse minima due to spot activity (Kalimeris et al. 2002) and/or even by the method of measuring the mid-eclipse times 
(Maceroni \& van't Veer 1994). The light-curves synthesis method developed by W-D is capable of extracting the conjunction instants 
and give more and better information with respect to the other methods, which do not consider spot activity and are based on 
the observations during minimum alone (Lee et al. 2009b). Because three datasets of NSVS 02502726 were modeled for spot parameters, 
we measured a minimum epoch for each eclipse in these datasets with the W-D code. Twenty-three light-curve timings and their errors 
are listed in Table 4, together with one ephemeris epoch (HJD 2453692.0280) presented in Table 1 of Coughlin \& Shaw (2007).
For ephemeris computations, weights were calculated as the inverse squares of the timing errors and were then scaled from 
the standard deviations ($\sigma$=0.00025 d) of all timing residuals. 

First of all, to obtain a mean light ephemeris for NSVS 02502726, we applied a linear least-squares fit to all eclipse timings
and found an improved ephemeris, as follows:
\begin{equation}
 C_1 = \mbox{HJD}~ 2,454,891.634172(34) + 0.559778168(30)E,
\end{equation}
where $E$ is the number of orbital cycles elapsed from the reference epoch and the parenthesized numbers are the 1$\sigma$-error values 
for the last digit of each term of the ephemeris. The resulting $O$--$C_{1}$ residuals calculated with equation (1) are listed in 
the fourth column of Table 4 and drawn in the upper part of Figure 5. As displayed in this figure, the orbital period of NSVS 02502726 
seemed to experience a parabolic variation. Therefore, by introducing all times of minimum light into a parabolic least-squares fit, 
we obtained the following quadratic ephemeris:
\begin{equation}
 C_2 = \mbox{HJD}~ 2,454,891.634581(91) + 0.559778460(68)E - 4.54(94) \times 10^{-10} E^2.
\end{equation}
In the upper part of Figure 5, the dashed curve represents the quadratic term of equation (2). The $O$--$C_{2}$ residuals from 
this ephemeris are given in the fifth column of Table 4. The quadratic ephemeris resulted in a smaller $\chi^2_{\rm red}$=1.91 
than the linear least-squares fit ($\chi^2_{\rm red}$=2.88).

The negative coefficient of the quadratic term indicates a continuous period decrease with a rate of 
d$P$/d$t$ = $-$5.9$\times$10$^{-7}$ d yr$^{-1}$. The period derivative corresponds to a fractional period change of 
$-$(1.62$\pm$0.34)$\times$10$^{-9}$, which is close to the rate of $-$(2.95$\pm$0.44)$\times$10$^{-9}$ derived from 
our W-D synthesis of all light and velocity curves (the unspotted model). Usually, such a variation in 
detached eclipsing binaries could be produced by angular momentum loss (AML) due to a magnetic stellar wind in the system. 
With the gyration constant $k^2$=0.1 typical for low-mass main sequence stars and with our absolute dimensions presented 
in the following section, the period decrease rate due to AML was calculated to be $-$4.3$\times$10$^{-9}$ d yr$^{-1}$ 
from the calibrated expression given by Guinan \& Bradstreet (1988). The value is two orders of magnitude too small to be 
the single cause of the observed period change. Thus, additional mechanism(s) may need to explain the secular change, 
most probably a light-travel-time (LTT) effect caused by the presence of a third object (cf. Lee et al. 2009a). 
It is also possible that the parabola of equation (2) could be only the observed part of a longer periodic variation. 

We fitted the minimum epochs to a sine curve instead of a quadratic term, as follows:
\begin{equation}
 C_3 = T_0 + PE + K \sin(\omega E + \omega_0 ). 
\end{equation}
The Levenberg-Marquart technique (Press et al. 1992) was used to yield the parameters given in Table 5, together with related 
quantities. The result is drawn at the upper part of Figure 5 with the solid curve. The timing residuals from the ephemeris (3) 
appear as $O$--$C_{3}$ in the sixth column of Table 4 and are plotted in the lower part of Figure 5. As can be seen in this figure, 
the entire collection of timings can be better fitted by a sine curve than by a quadratic ephemeris but a large number of 
future accurate timings is required before this can be tested at an acceptable level. If the sinusoidal variation represents 
a real period change in NSVS 02502726, it most likely arises from the LTT effect driven by the existence of a third component 
orbiting the eclipsing binary. The LTT orbit has a period of 2.51 yrs, a semi-amplitude of 0.0011 d, and 
a projected orbital semi-major axis of 0.19 au. The mass function of this object becomes $f(M_{3})$=0.00116 $M_\odot$ and 
its minimum mass is $M_3 \sin i_3$=0.11 $M_\odot$. The hypothetical third body would be difficult to detect photometrically 
and spectroscopically, because it contributes only 0.5\% to the total light of the triple system. More systematic and 
continuous observations of eclipse timings are required to identify and understand the orbital period change.

\section{SUMMARY AND DISCUSSION}

New multiband CCD light curves plus the previous CIG's data all display complete eclipses leading to well-determined system parameters. 
The light curves of NSVS 02502726 are best fitted by using a two-spot model with one cool spot on each component and 
the seasonal light variability has been ascribed to changes in the spot parameters with time. These spots may be formed from
magnetic dynamo-related activity because the system is rotating rapidly and the component stars have a deep convective envelope. 
The first period study of the eclipsing pair indicates a continuous period decrease at a rate of $-$5.9$\times$10$^{-7}$ d yr$^{-1}$
or a sinusoidal variation with a period and semi-amplitude of 2.51 yrs and 0.0011 d, respectively. We speculate that 
the orbital period change has varied due to the existence of a gravitationally-bound low-mass tertiary companion overlaid on AML 
via magnetic stellar wind braking, rather than in a single cause.

The light and velocity solutions allow us to compute the physical properties of NSVS 02502726. As listed in Table 3, there exist 
discrepancies between the binary parameters derived from the three datasets, which might be mainly caused by the effect of 
spots. We adopted a weighted average of those parameters as our final values and thus determined the absolute parameters 
for each component of the system listed in Table 6, together with those of CIG for comparison. 
The luminosity ($L$) and bolometric magnitudes ($M_{\rm bol}$) were derived by using $T_{\rm eff}$$_\odot$=5,780 K and 
$M_{\rm bol}$$_\odot$=+4.73 for solar values. It was assumed that the temperature of each component has an error of 200 K as 
given by CIG, because the temperature error listed in Table 3 is certainly an underestimate. The bolometric corrections were 
obtained from the relation between $\log T_{\rm eff}$ and BC given by Torres (2010). With an apparent visual magnitude of 
$V$=+13.41 at maximum light and the interstellar absorption of $A_{\rm V}$=0.07, we calculated a distance to the system 
of 163$\pm$15 pc. In this paper, we determined the masses and radii of NSVS 02502726 with an accuracy of about 3\% and 1\%, 
respectively. Our values of the masses are consistent with those of CIG within the limits of their errors, but the radii are 
very far from each other. Further, our result represents that the primary star is larger than the secondary, which is reversed 
for that of CIG. At present, we cannot offer the correct explanation for this discrepancy but the possibility is that 
the $RI$ light curves analyzed by CIG have a relatively large amount of light changes.

Using the physical parameters of NSVS 02502726, we examined the evolutionary state of the system in mass-radius and 
mass-temperature diagrams. The locations of the component stars in these diagrams are shown in Figure 6, together with 
the other well-studied detached eclipsing binaries in the range of 0.2$-$0.8 $M_\odot$. The data are taken from 
Bayless \& Orosz (2006) and DEBCat\footnote {http://www.astro.keele.ac.uk/$\sim$jkt/debcat/}. 
We compared the physical parameters to the predicted values from the stellar evolutionary models of Baraffe et al. (1998) 
for solar metallicity. In the diagrams, the primary star of NSVS 02502726 has a best-fit model when the 79-Myr isochrone is 
used, which means that the eclipsing system has not yet reached the main sequence. On the contrary, the secondary component 
is by far larger and hotter than expected for its mass in the same age. We also show in Figure 6 the empirical mass-radius 
relation calibrated by Bayless \& Orosz (2006) and the 300 Myr isochrones from the Dartmouth series (Dotter et al. 2008) 
for comparison. Even in this empirical case, the parameters of the primary are in good agreement with the prediction, 
but the radius of the secondary star is oversized by about 85\%. The inflated radius might result from the effect of 
enhanced magnetic activity in the young secondary with a mass of about 0.35 $M_\odot$, the boundary at which stars are 
thought to become fully convective (Chabrier \& Baraffe 1997). Then, the effective temperature predicted from models does 
depend strongly on metallicity. Further spectroscopy, photometry, and times of minimum light are needed to determine 
the metallicity and to resolve the current discrepancy with the models and the orbital period variation of the system. 
These would clearly make NSVS 02502726 even more interesting.

\acknowledgments{ }
We would like to thank \" Om\"ur \c Cakirli for sending us their data on NSVS 02502726 and the staffs of LOAO for assistance 
with our observations. We appreciate the careful reading and valuable comments of the anonymous referee. This research has made 
use of the Simbad database maintained at CDS, Strasbourg, France. This work was supported by the KASI 
(Korea Astronomy and Space Science Institute) grant 2012-1-410-02.

\clearpage
\begin{figure}
 \includegraphics[]{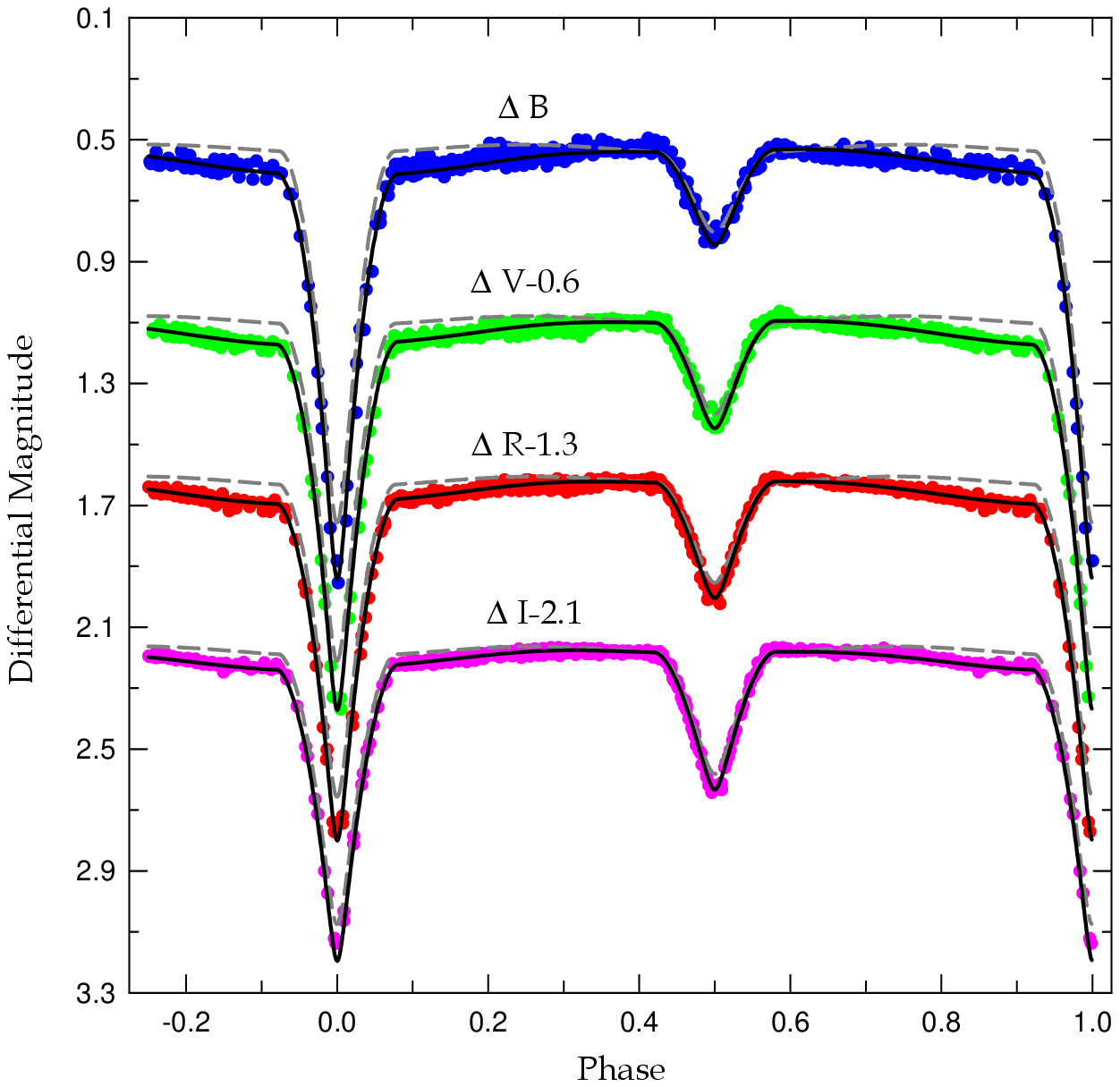}
 \caption{$BVRI$ light curves of NSVS 02502726 observed in 2009 with the fitted models. The circles are individual measures 
 and the dashed and solid lines represent the synthetic curves obtained from no spot and the two-spot model, respectively. }
 \label{Fig1}
\end{figure}

\begin{figure}
 \includegraphics[]{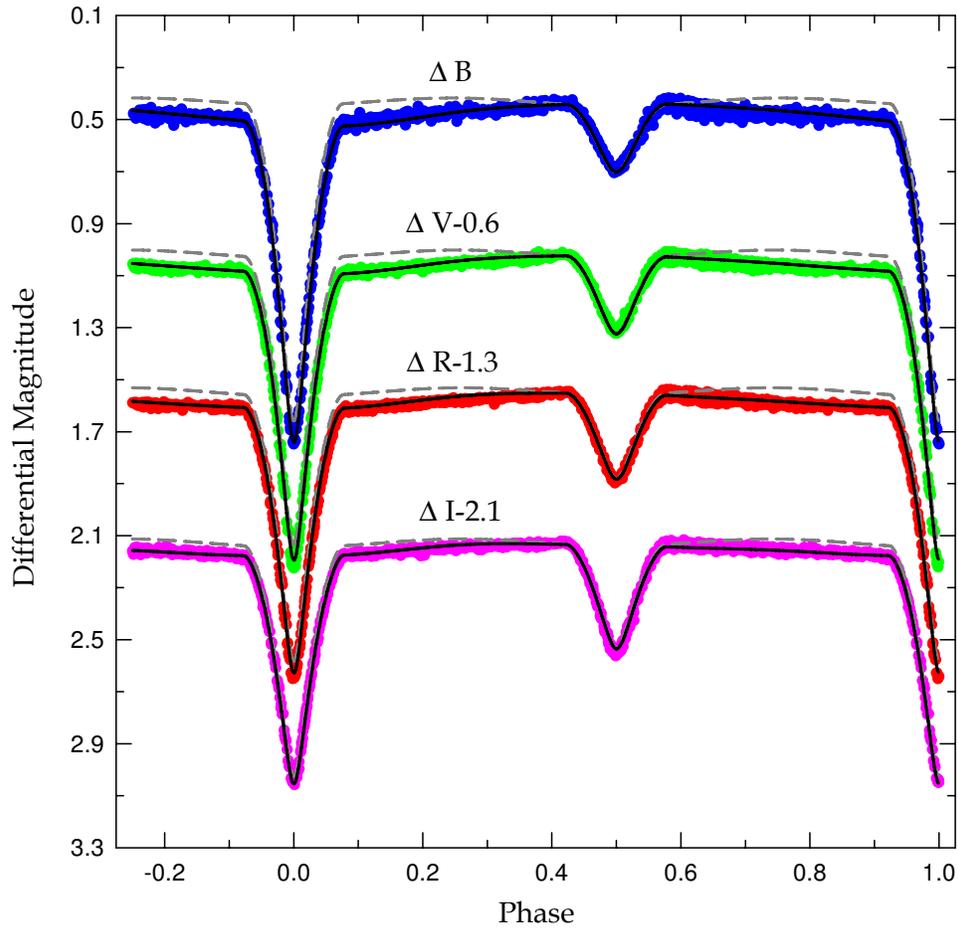}
  \caption{In the same sense as Figure 1, $BVRI$ light curves of NSVS 02502726 observed in 2011. }
\label{Fig2}
\end{figure}

\begin{figure}
 \includegraphics[]{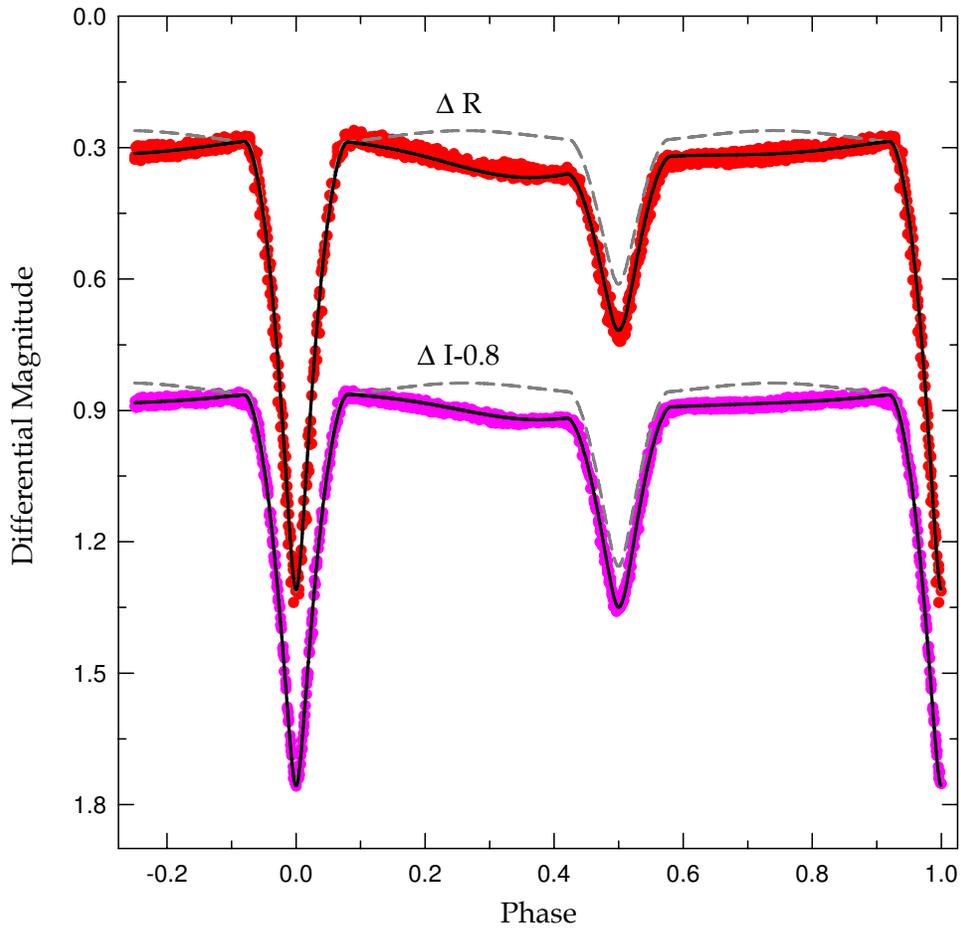}
 \caption{In the same sense as Figure 1, $RI$ light curves of NSVS 02502726 observed by \c Cakirli et al. (2009) in 2008. } 
\label{Fig3}
\end{figure}

\begin{figure}
 \includegraphics[]{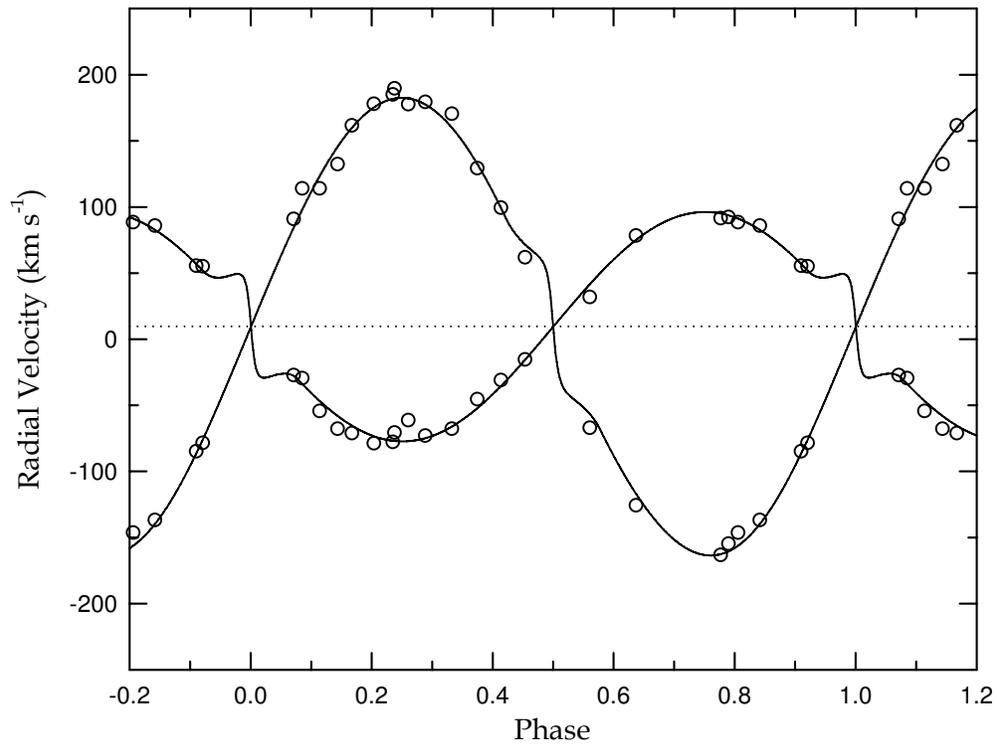}
 \caption{Radial-velocity curves of NSVS 02502726. The open circles are the measures of \c Cakirli et al. (2009), while 
 the continuous curves denote the result from consistent light and velocity curve analysis including proximity effects listed 
 in Table 6. The dotted line refers to the system velocity of 9.7 km s$^{-1}$. }
 \label{Fig4}
\end{figure}

\begin{figure}
 \includegraphics[]{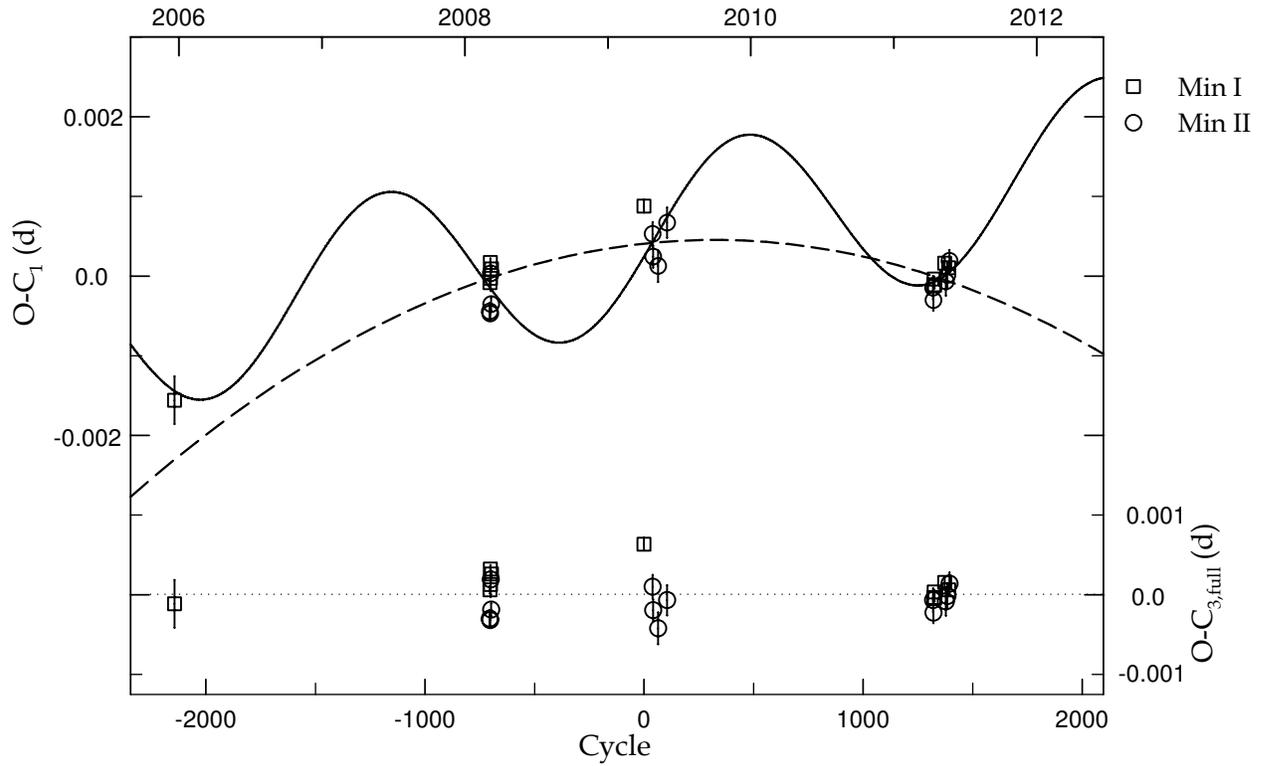}
 \caption{$O$--$C$ diagram of NSVS 02502726 with respect to the linear ephemeris (1). In the upper part, the dashed parabola is 
 the quadratic fit to all times of minimum light and the solid curve represents the result obtained by fitting a sine wave to
 these timings. The residuals from the ephemeris (3) are plotted in the lower part.  }
 \label{Fig5}
\end{figure}

\begin{figure}
 \includegraphics[]{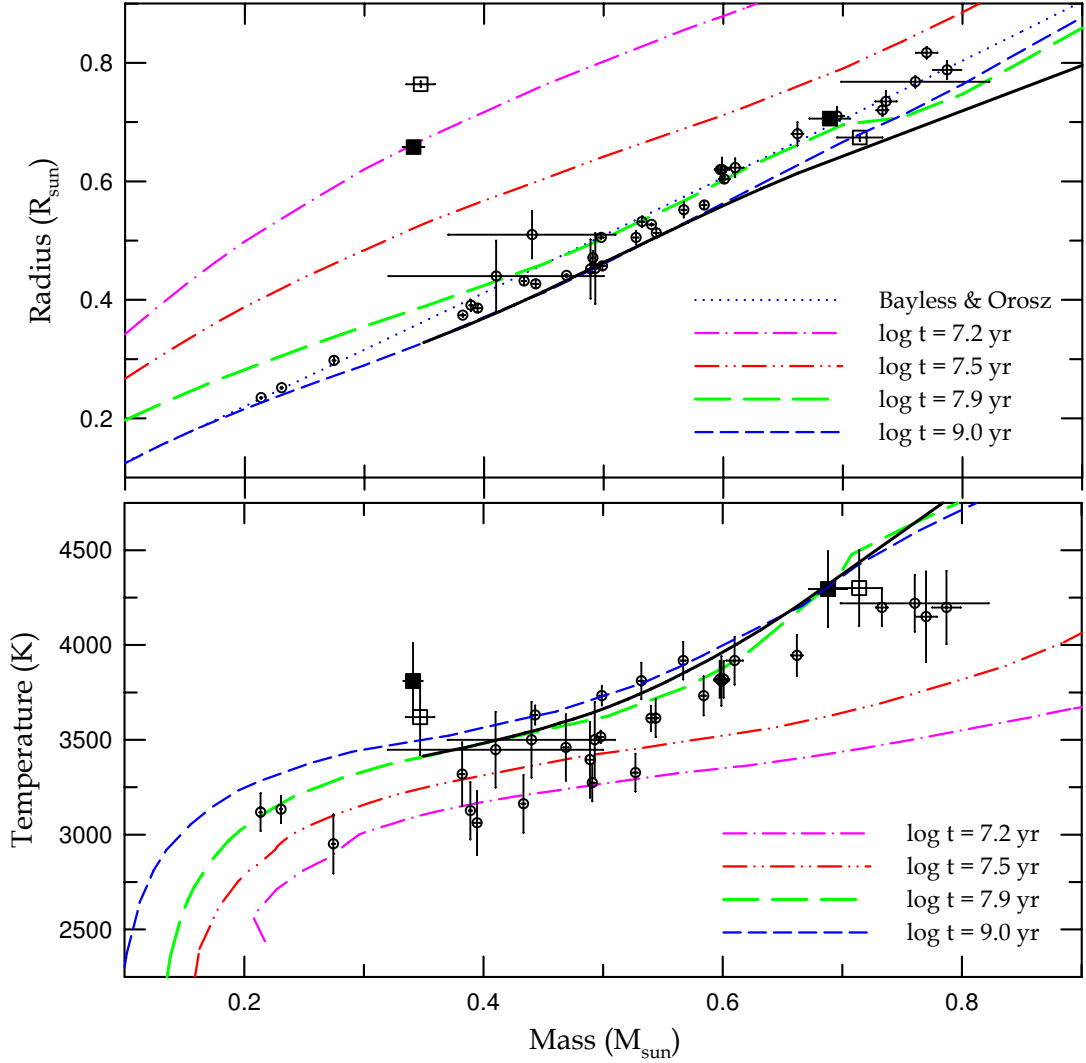}
 \caption{Mass-radius (upper) and mass-temperature (lower) diagrams of the stars between 0.2 and 0.8 $M_\odot$ in double-lined, 
 detached eclipsing binaries. The dashed lines correspond to the isochrones from the Baraffe et al. (1998) models for $\log$ ages 
 = 7.2, 7.5, 7.9, and 9.0 yr, respectively. Overlaid are solar-metallicity Dartmouth isochrones at 300 Myr with the solid lines 
 for comparison. In the upper panel, the dotted line represents the empirical mass-radius relation of Bayless \& Orosz (2006). 
 Our results for NSVS 02502726 are plotted as the filled squares, while those of \c Cakirli et al. (2009) shown as the open squares. }
\label{Fig6}
\end{figure}

\clearpage
\begin{deluxetable}{crcrcrcr}
\tabletypesize{\small}
\tablewidth{0pt} 
\tablecaption{CCD photometric data of NSVS 02502726 observed in 2009 and 2011.}
\tablehead{
\colhead{HJD} & \colhead{$\Delta B$} & \colhead{HJD} & \colhead{$\Delta V$} & \colhead{HJD} & \colhead{$\Delta R$} & \colhead{HJD} & \colhead{$\Delta I$} 
}                                                                                                                  
\startdata                                                                                                         
2,454,891.60180  &  0.680   &   2,454,891.60944  &  0.814   &   2,454,891.61079  &  0.661   &   2,454,891.61173  &  0.392     \\
2,454,891.61377  &  0.978   &   2,454,891.61628  &  1.017   &   2,454,891.61777  &  0.865   &   2,454,891.61876  &  0.565     \\
2,454,891.62083  &  1.263   &   2,454,891.62334  &  1.280   &   2,454,891.62483  &  1.129   &   2,454,891.62586  &  0.801     \\
2,454,891.62796  &  1.607   &   2,454,891.63047  &  1.629   &   2,454,891.63200  &  1.440   &   2,454,891.63305  &  1.022     \\
2,454,891.63515  &  1.883   &   2,454,891.63766  &  1.733   &   2,454,891.63916  &  1.420   &   2,454,891.64022  &  0.933     \\
2,454,891.64235  &  1.635   &   2,454,891.64485  &  1.402   &   2,454,891.64635  &  1.093   &   2,454,891.64740  &  0.686     \\
2,454,891.64953  &  1.234   &   2,454,891.65204  &  1.092   &   2,454,891.65353  &  0.830   &   2,454,891.65459  &  0.480     \\
2,454,891.65672  &  0.992   &   2,454,891.65923  &  0.857   &   2,454,891.66072  &  0.626   &   2,454,891.66178  &  0.322     \\
2,454,891.66391  &  0.777   &   2,454,891.66642  &  0.681   &   2,454,891.66791  &  0.486   &   2,454,891.66897  &  0.191     \\
2,454,891.67110  &  0.655   &   2,454,891.67361  &  0.577   &   2,454,891.67510  &  0.386   &   2,454,891.67616  &  0.121     \\
\enddata
\tablecomments{This table is available in its entirety in machine-readable and Virtual Observatory (VO) forms 
in the online journal. A portion is shown here for guidance regarding its form and content.}
\end{deluxetable}

\begin{deluxetable}{lccc}
\tablewidth{0pt}
\tablecaption{Radial velocity and light-curve sets for NSVS 02502726.}
\tablehead{
\colhead{Reference}          & \colhead{Season} & \colhead{Data type} & \colhead{$\sigma$$\rm ^a$} }
\startdata
\c Cakirli et al. (2009)     & 2008           & RV1                 & 5.7 km s$^{-1}$   \\
                             &                & RV2                 & 6.8 km s$^{-1}$   \\
                             & 2008           & $R$                 & 0.0124            \\
                             &                & $I$                 & 0.0073            \\
LOAO                         & 2009           & $B$                 & 0.0194            \\
                             &                & $V$                 & 0.0152            \\
                             &                & $R$                 & 0.0147            \\
                             &                & $I$                 & 0.0117            \\
                             & 2011           & $B$                 & 0.0136            \\
                             &                & $V$                 & 0.0096            \\
                             &                & $R$                 & 0.0081            \\
                             &                & $I$                 & 0.0078            \\
\enddata
\tablenotetext{a}{For the light curves, in units of total light at orbital phases of light maxima.}
\end{deluxetable}

\begin{deluxetable}{lcccccccc}
\tabletypesize{\scriptsize}
\tablewidth{0pt} 
\tablecaption{Velocity and light curve parameters of NSVS 02502726 calculated from the W-D code.}
\tablehead{
\colhead{Parameter}          & \multicolumn{2}{c}{RV+$\rm LC_{LOAO2009}$} && \multicolumn{2}{c}{RV+$\rm LC_{LOAO2011}$} && \multicolumn{2}{c}{RV+$\rm LC_{CIG2008}$}    \\ [1.0mm] \cline{2-3} \cline{5-6} \cline{8-9} \\[-2.0ex]
                             & \colhead{Primary} & \colhead{Secondary}    && \colhead{Primary} & \colhead{Secondary}    && \colhead{Primary} & \colhead{Secondary}              
}
\startdata                                                                                
$T_0$ (HJD)$\rm ^a$          & \multicolumn{2}{c}{4,891.635203(51)}       && \multicolumn{2}{c}{5,631.660891(21)}       && \multicolumn{2}{c}{4,496.990484(44)}         \\          
$P$ (d)                      & \multicolumn{2}{c}{0.55977301(96)}         && \multicolumn{2}{c}{0.55977952(40)}         && \multicolumn{2}{c}{0.55979508(862)}          \\          
$\gamma$ (km s$^{-1}$)       & \multicolumn{2}{c}{9.76(83)}               && \multicolumn{2}{c}{9.58(1.08)}             && \multicolumn{2}{c}{9.76(1.14)}               \\          
$a$ (R$_\odot$)              & \multicolumn{2}{c}{2.887(23)}              && \multicolumn{2}{c}{2.882(31)}              && \multicolumn{2}{c}{2.887(32)}                \\          
$q$                          & \multicolumn{2}{c}{0.4956(58)}             && \multicolumn{2}{c}{0.4956(27)}             && \multicolumn{2}{c}{0.4956(32)}               \\          
$i$ ($^\circ$)               & \multicolumn{2}{c}{88.00(13)}              && \multicolumn{2}{c}{86.71(4)}               && \multicolumn{2}{c}{87.21(6)}                 \\          
$T$ (K)                      & 4,298(38)            & 3,843(25)           && 4,296(15)            & 3,789(10)           && 4,293(19)            & 3,843(13)             \\          
$\Omega$                     & 4.573(22)            & 3.419(26)           && 4.648(11)            & 3.474(11)           && 4.539(15)            & 3.395(12)             \\          
$\Omega_{\rm in}$$\rm ^b$    & \multicolumn{2}{c}{2.867}                  && \multicolumn{2}{c}{2.867}                  && \multicolumn{2}{c}{2.867}                    \\
$A$                          & 0.5                  & 0.5                 && 0.5                  & 0.5                 && 0.5                  & 0.5                   \\          
$g$                          & 0.32                 & 0.32                && 0.32                 & 0.32                && 0.32                 & 0.32                  \\          
$X$, $Y$                     & 0.245, 0.430         & 0.042, 0.591        && 0.244, 0.431         & 0.008, 0.617        && 0.244, 0.432         & 0.042, 0.591          \\          
$x_{B}$, $y_{B}$             & 1.139, -0.354        & 0.655, 0.204        && 1.138, -0.354        & 0.552, 0.326        && \dots                & \dots                 \\          
$x_{V}$, $y_{V}$             & 0.754,  0.051        & 0.441, 0.420        && 0.753,  0.051        & 0.376, 0.498        && \dots                & \dots                 \\          
$x_{R}$, $y_{R}$             & 0.466,  0.317        & 0.307, 0.504        && 0.466,  0.317        & 0.271, 0.545        && 0.466,  0.317        & 0.307, 0.504          \\          
$x_{I}$, $y_{I}$             & 0.262,  0.440        & 0.079, 0.646        && 0.261,  0.440        & 0.044, 0.686        && 0.261,  0.441        & 0.079, 0.646          \\          
$L/(L_1+L_2)_{B}$            & 0.7580(33)           & 0.2420              && 0.7823(15)           & 0.2177              && \dots                & \dots                 \\          
$L/(L_1+L_2)_{V}$            & 0.7277(33)           & 0.2723              && 0.7518(14)           & 0.2482              && \dots                & \dots                 \\          
$L/(L_1+L_2)_{R}$            & 0.7039(35)           & 0.2961              && 0.7270(15)           & 0.2730              && 0.7013(18)           & 0.2987                \\          
$L/(L_1+L_2)_{I}$            & 0.6560(37)           & 0.3440              && 0.6739(17)           & 0.3261              && 0.6535(23)           & 0.3465                \\          
$r$ (pole)                   & 0.2444(13)           & 0.2262(28)          && 0.2400(6)            & 0.2207(12)          && 0.2464(9)            & 0.2286(14)            \\          
$r$ (point)                  & 0.2507(15)           & 0.2423(39)          && 0.2458(7)            & 0.2351(16)          && 0.2530(10)           & 0.2456(20)            \\          
$r$ (side)                   & 0.2472(14)           & 0.2302(30)          && 0.2426(7)            & 0.2244(13)          && 0.2493(9)            & 0.2328(15)            \\          
$r$ (back)                   & 0.2496(15)           & 0.2385(36)          && 0.2448(7)            & 0.2318(15)          && 0.2518(10)           & 0.2415(18)            \\
$r$ (volume)$\rm ^c$         & 0.2472(14)           & 0.2318(33)          && 0.2425(7)            & 0.2258(14)          && 0.2493(10)           & 0.2345(33)            \\ [1.0mm]  
\multicolumn{9}{l}{Spot parameters:}                                                                                                                                    \\          
Colatitude (deg)             & 25.1                 & 91.4                && 17.5                 & 96.5                && 34.5                 & 77.0                  \\          
Longitude (deg)              & 320.5                & 243.2               && 327.1                & 272.1               && 241.6                & 276.4                 \\          
Radius (deg)                 & 49.4                 & 33.5                && 64.1                 & 35.2                && 36.2                 & 37.0                  \\          
Temp. factor                 & 0.95                 & 0.92                && 0.95                 & 0.93                && 0.82                 & 0.89                  \\ [1.0mm]
\multicolumn{9}{l}{Absolute parameters:}                                                                                                                                \\
$M$($M_\odot$)               & 0.690(16)            & 0.342(9)            && 0.686(16)            & 0.340(8)            && 0.689(17)            & 0.342(9)              \\
$R$($R_\odot$)               & 0.713(8)             & 0.669(10)           && 0.698(7)             & 0.650(8)            && 0.719(8)             & 0.676(10)             \\
\enddata
\tablenotetext{a}{HJD 2,450,000 is suppressed.}
\tablenotetext{b}{Potential for the inner critical surface.}
\tablenotetext{c}{Mean volume radius.}
\end{deluxetable}

\begin{deluxetable}{llrccccl}
\tabletypesize{\small} 
\tablewidth{0pt}
\tablecaption{CCD timings of minimum light for NSVS 02502726.}
\tablehead{
\colhead{HJD} & \colhead{Error} & \colhead{Epoch} & \colhead{$O$--$C_1$} & \colhead{$O$--$C_2$} & \colhead{$O$--$C_3$} & \colhead{Min} & References \\
\colhead{(2,450,000+)} &  & & & & & &  }                                                               
\startdata
3,692.0280   &  $\pm$0.0003     & $-$2143.0        &  $-$0.00156  &  $+$0.00075  &  $-$0.00011  &  II  &  Coughlin \& Shaw (2007)       \\  
4,497.27001  &  $\pm$0.00011    &  $-$704.5        &  $-$0.00044  &  $-$0.00042  &  $-$0.00030  &  II  &  \c Cakirli et al. (2009)      \\  
4,497.55026  &  $\pm$0.00005    &  $-$704.0        &  $-$0.00008  &  $-$0.00006  &  $+$0.00006  &  I   &  \c Cakirli et al. (2009)      \\  
4,498.38954  &  $\pm$0.00007    &  $-$702.5        &  $-$0.00047  &  $-$0.00045  &  $-$0.00032  &  II  &  \c Cakirli et al. (2009)      \\  
4,498.67007  &  $\pm$0.00005    &  $-$702.0        &  $+$0.00017  &  $+$0.00019  &  $+$0.00032  &  I   &  \c Cakirli et al. (2009)      \\  
4,499.22965  &  $\pm$0.00016    &  $-$701.0        &  $-$0.00003  &  $-$0.00001  &  $+$0.00012  &  I   &  \c Cakirli et al. (2009)      \\  
4,499.50960  &  $\pm$0.00010    &  $-$700.5        &  $+$0.00003  &  $+$0.00005  &  $+$0.00019  &  II  &  \c Cakirli et al. (2009)      \\  
4,500.62877  &  $\pm$0.00008    &  $-$698.5        &  $-$0.00035  &  $-$0.00033  &  $-$0.00018  &  II  &  \c Cakirli et al. (2009)      \\  
4,501.46888  &  $\pm$0.00004    &  $-$697.0        &  $+$0.00009  &  $+$0.00011  &  $+$0.00026  &  I   &  \c Cakirli et al. (2009)      \\  
4,891.63505  &  $\pm$0.00008    &       0.0        &  $+$0.00088  &  $+$0.00047  &  $+$0.00063  &  I   &  This work (LOAO 2009)         \\  
4,913.74594  &  $\pm$0.00015    &      39.5        &  $+$0.00053  &  $+$0.00011  &  $+$0.00010  &  II  &  This work (LOAO 2009)         \\  
4,914.86521  &  $\pm$0.00014    &      41.5        &  $+$0.00024  &  $-$0.00018  &  $-$0.00019  &  II  &  This work (LOAO 2009)         \\  
4,927.73999  &  $\pm$0.00020    &      64.5        &  $+$0.00013  &  $-$0.00030  &  $-$0.00042  &  II  &  This work (LOAO 2009)         \\  
4,950.69144  &  $\pm$0.00019    &     105.5        &  $+$0.00067  &  $+$0.00024  &  $-$0.00006  &  II  &  This work (LOAO 2009)         \\  
5,629.70154  &  $\pm$0.00015    &    1318.5        &  $-$0.00015  &  $-$0.00015  &  $-$0.00006  &  II  &  This work (LOAO 2011)         \\  
5,630.82094  &  $\pm$0.00013    &    1320.5        &  $-$0.00030  &  $-$0.00030  &  $-$0.00022  &  II  &  This work (LOAO 2011)         \\  
5,631.66080  &  $\pm$0.00004    &    1322.0        &  $-$0.00011  &  $-$0.00011  &  $-$0.00003  &  I   &  This work (LOAO 2011)         \\  
5,632.78043  &  $\pm$0.00003    &    1324.0        &  $-$0.00004  &  $-$0.00004  &  $+$0.00003  &  I   &  This work (LOAO 2011)         \\  
5,633.89991  &  $\pm$0.00004    &    1326.0        &  $-$0.00011  &  $-$0.00011  &  $-$0.00004  &  I   &  This work (LOAO 2011)         \\  
5,659.64998  &  $\pm$0.00006    &    1372.0        &  $+$0.00016  &  $+$0.00021  &  $+$0.00015  &  I   &  This work (LOAO 2011)         \\  
5,662.72853  &  $\pm$0.00018    &    1377.5        &  $-$0.00007  &  $-$0.00002  &  $-$0.00008  &  II  &  This work (LOAO 2011)         \\  
5,666.64706  &  $\pm$0.00016    &    1384.5        &  $+$0.00001  &  $+$0.00007  &  $-$0.00001  &  II  &  This work (LOAO 2011)         \\  
5,669.72593  &  $\pm$0.00004    &    1390.0        &  $+$0.00010  &  $+$0.00017  &  $+$0.00006  &  I   &  This work (LOAO 2011)         \\  
5,671.68524  &  $\pm$0.00014    &    1393.5        &  $+$0.00019  &  $+$0.00026  &  $+$0.00013  &  II  &  This work (LOAO 2011)         \\  
\enddata
\end{deluxetable}

\begin{deluxetable}{lcc}
\tablewidth{0pt}
\tablecaption{The fitted parameters for ephemeris (3) of NSVS 02502726. }
\tablehead{
\colhead{Parameter}     &  \colhead{Value}                &  \colhead{Unit}
}                                                         
\startdata                                                
$T_0$                   &  2,454,891.63462$\pm$0.00010    &  HJD             \\
$P$                     &  0.559778606$\pm$0.000000095    &  d               \\
$K$                     &  0.00112$\pm$0.00015            &  d               \\
$\omega$                &  0.00384$\pm$0.00017            &  rad/P           \\
$\omega$$_0$            &  6.10$\pm$0.12                  &  rad             \\[1.0mm]
$P_{3}$                 &  2.510$\pm$0.062                &  yr              \\
$a_{12}\sin i_{3}$      &  0.194$\pm$0.027                &  au              \\
$f(M_{3})$              &  0.00116$\pm$0.00016            &  $M_\odot$       \\       
$M_3 \sin i_{3}$        &  0.115$\pm$0.010                &  $M_\odot$       \\
$\chi^2 _{\rm red}$     &  1.027                          &                  \\
\enddata
\end{deluxetable}

\begin{deluxetable}{lccccc}
\tablewidth{0pt} 
\tablecaption{Physical properties of NSVS 02502726$\rm ^a$.}
\tablehead{
\colhead{Parameter}          & \multicolumn{2}{c}{Cakirli et al.}         && \multicolumn{2}{c}{This Work}               \\ [1.0mm] \cline{2-3} \cline{5-6} \\[-2.0ex]
                             & \colhead{Primary} & \colhead{Secondary}    && \colhead{Primary} & \colhead{Secondary}            
}
\startdata 
$\gamma$ (km s$^{-1}$)       & \multicolumn{2}{c}{3.15(44)}               && \multicolumn{2}{c}{9.71(98)}                \\          
$a$ (R$_\odot$)              & \multicolumn{2}{c}{2.914(26)}              && \multicolumn{2}{c}{2.886(27)}               \\          
$q$                          & \multicolumn{2}{c}{0.486(10)}              && \multicolumn{2}{c}{0.4956(32)}              \\          
$i$ ($^\circ$)               & \multicolumn{2}{c}{87.0(1.0)}              && \multicolumn{2}{c}{86.93(5)}                \\          
$T$ (K)                      & 4,300(200)           & 3,620(205)          && 4,295(200)$\rm ^b$   & 3,812(200)$\rm ^b$   \\          
$\Omega$                     & 4.804(2)             & 3.862(3)            && 4.602(13)            & 3.436(13)            \\          
$r$                          & 0.2312(5)            & 0.2619(8)           && 0.2451(9)            & 0.2278(19)           \\ [1.0mm] 
$M$ ($M_\odot$)              & 0.714(19)            & 0.347(12)           && 0.689(16)            & 0.341(9)             \\          
$R$ ($R_\odot$)              & 0.674(6)             & 0.763(7)            && 0.707(7)             & 0.657(8)             \\          
$\log$ $g$ (cgs)             & 4.635(4)             & 4.213(8)            && 4.578(14)            & 4.336(16)            \\          
$\log$ $\rho$ ($\rho_\odot$) & \dots                & \dots               && 1.954(77)            & 1.206(55)            \\          
$L$ ($L_\odot$)              & 0.139(14)            & 0.090(10)           && 0.152(29)            & 0.082(17)            \\          
$M_{\rm bol}$ (mag)          & $+$6.88(14)          & $+$7.35(12)         && $+$6.77(20)          & $+$7.45(23)          \\          
BC (mag)                     & \dots                & \dots               && $-$0.77              & $-$1.46              \\          
$M_{V}$ (mag)                & $+$7.73(13)          & $+$9.11(14)         && $+$7.54(20)          & $+$8.91(23)          \\          
Distance (pc)                & \multicolumn{2}{c}{173(8)}                 && \multicolumn{2}{c}{163(15)}                 \\          
\enddata
\tablenotetext{a}{Parameters correspond to the weighted mean of the results in Table 3.}
\tablenotetext{b}{Errors are assigned to a value given by Cakirli et al. (2009).}
\end{deluxetable}

\end{document}